\documentclass[pmlr,twocolumn,10pt]{jmlr} 
\mlhtrack{proceedings}

\newif\iffinal
\finaltrue  

\usepackage{soul}
\usepackage{booktabs}
\usepackage{siunitx}
\usepackage{multirow}
\usepackage{makecell}
\usepackage[switch]{lineno}
\usepackage{pifont}
\usepackage{adjustbox}
\usepackage{xcolor}
\newcommand{\cmark}{\textcolor{green!80!black}{\ding{51}}}
\newcommand{\xmark}{\textcolor{red!90!black}{\ding{55}}}
\usepackage{stfloats}
\usepackage{arydshln}
\usepackage[table]{xcolor}

\theorembodyfont{\upshape}
\theoremheaderfont{\scshape}
\theorempostheader{:}
\theoremsep{\newline}

\newcommand{\modelname}{CoRE}
\title[CoRE]{CoRE: Relating Fine-Grained Chest X-ray Findings in Anatomical Context}

\author{Anonymous Author(s)}

\author{%
 \Name{Minjun Kim} \Email{jayden.kim@yejix.ai}\\
 \Name{Jong Hak Moon}\footnotemark[2] \Email{jh.moon@yejix.ai}\\
 \addr Yeji X, Seoul, South Korea
}

\begin{document}

\maketitle
\renewcommand{\thefootnote}{\fnsymbol{footnote}} 
\footnotetext[2]{Corresponding author.}
\renewcommand{\thefootnote}{\arabic{footnote}} %
\ifmlhdemo\else
\begin{abstract}
Chest X-ray interpretation depends on both the anatomical location of findings and their relationships with other abnormalities. However, relations defined over global finding labels do not explicitly distinguish the same finding across anatomical locations.
We propose \modelname{} (Co-occurrence-guided Region Evidence), a framework that incorporates relationships among findings and anatomical context into fine-grained chest X-ray prediction. It groups anatomical region–finding labels by co-occurrence and uses these groups to guide shared representation learning through visual prompt tuning on a frozen vision transformer. 
The predictions are supplemented by an anatomically supervised region-evidence stream through region-matched logit residuals. Experiments on MIMIC-CXR with Chest ImaGenome annotations show that CoRE improves region–finding classification over standard and relation-aware prompt-based baselines. Quantitative evaluations on MS-CXR and the Chest ImaGenome gold standard show improved finding localization, supported by qualitative comparisons. CoRE also improves finding-level classification on mapped finding labels in external VinDr-CXR and CheXpert datasets without target-domain adaptation. These results demonstrate the value of jointly modeling finding relationships and anatomical context for fine-grained chest X-ray analysis.

\end{abstract}
\fi

\ifmlhneedsstatements
\paragraph*{Data and Code Availability}
All datasets used in this study are publicly accessible, with details
provided in Section~\ref{sec:exp}. Our implementation, including preprocessing and
training code, is available at: https://github.com/yejix-ai/CoRE.

\paragraph*{Institutional Review Board (IRB)}
IRB approval was not required because this study used only de-identified, previously released datasets.
\fi

\section{Introduction}
\label{sec:intro}

Chest X-ray (CXR) findings are interpreted in anatomical context.
Combinations of abnormalities may provide complementary diagnostic evidence; for example, cardiomegaly, pulmonary vascular redistribution, and pleural effusion may collectively support a diagnosis of congestive heart failure~\citep{gluecker1999,chen1992,cardinale2014,rabaey2025modeling}.
The diagnostic implications of a finding can also depend on its anatomical location: upper-lobe opacity may suggest post-primary tuberculosis or granulomatous disease, whereas dependent opacity may suggest aspiration pneumonia~\citep{bhardwaj2023radiographic,jeong2008pulmonary,franquet2000aspiration,dibardino2015aspiration}.
These examples illustrate that radiographic interpretation depends not only on what a finding is, but also on where it occurs and which other abnormalities accompany it, motivating models that jointly account for finding identity, anatomical location, and relationships among findings.

Label-correlation methods in multi-label image recognition use co-occurrence statistics or graph-based relations to model dependencies among labels \citep{chen2019mlgcn,ma2025mlvpt}, and related approaches have also been developed for multi-label CXR classification \citep{chen2020chexgcn,wang2023bbgcn}.
However, these relations are typically defined over global finding labels and therefore do not distinguish the same finding across anatomical locations.

A separate line of work uses localized visual evidence for CXR prediction and localization.
Anatomy-aware prediction and localized detection methods use anatomical region annotations or detection models to associate finding predictions with region-level evidence \citep{wu2021imagenome_anaxnet,muller2023adpd,dehaghani2024chexdetector}.
Vision-language and phrase-grounding methods instead align radiology text, textual prompts, or diagnostic phrases with image regions, providing localized visual evidence for queried concepts \citep{boecking2022biovil,mueller2024chex,medgrounder2025,zou2025umedground}.
These methods associate individual findings with anatomical locations or provide localized visual evidence, but do not typically treat anatomically specified findings as the units for defining label relations or co-occurrence structure. This motivates using region–finding pairs as a common unit for both prediction and co-occurrence-guided label representation, preserving anatomical specificity in both the prediction targets and the organization of label representations.

To address this gap, we propose \modelname{} (\textbf{Co}-occurrence-guided \textbf{R}egion \textbf{E}vidence), a framework for fine-grained CXR multi-label prediction.
Using anatomical region--finding pairs (e.g., left upper lung zone--pneumonia) as prediction targets, \modelname{} uses co-occurrence structure among these anatomically specified labels to organize label representations and incorporates region-matched visual evidence into each prediction.
Together, these components preserve anatomical specificity while incorporating co-occurrence context, reflecting the contextual nature of CXR interpretation.
\section{Related Work}
\label{sec:rw}

\paragraph{Label Correlation Modeling for Multi-label Prediction.}

Multi-label image classification has explored a variety of approaches for exploiting dependencies among labels
~\citep{wang2016cnnrnn,lanchantin2021ctran,liu2021query2label}. ML-GCN~\citep{chen2019mlgcn} constructs a label graph from conditional co-occurrence statistics and learns label-specific classifiers through graph propagation, while ML-VPT~\citep{ma2025mlvpt} incorporates correlative and discriminative label relations through two parallel prompt-based prediction pathways built on a frozen vision transformer. In CXR classification, CheXGCN~\citep{chen2020chexgcn} uses pathology co-occurrence to recalibrate image-based predictions, GL-MLL~\citep{li2022glmll} models label correlations from globally static and locally adaptive views, and BB-GCN~\citep{wang2023bbgcn} models global pathology co-occurrence with a graph convolution network and couples the resulting label representations with image features. These approaches typically model relations among global finding labels, without explicitly incorporating anatomical specificity into the relation unit.

\paragraph{Anatomy-aware CXR Prediction.}
Prior CXR models have used anatomical regions to localize abnormalities and support region-specific prediction.
ADPD~\citep{muller2023adpd} uses anatomical region boxes as spatial proxies for pathology localization and predicts pathologies within individual regions.
CheXDetector~\citep{dehaghani2024chexdetector} detects anatomical regions and extracts localized visual representations for region-wise disease prediction.
AnaXNet~\citep{wu2021imagenome_anaxnet} represents detected anatomical regions as graph nodes and aggregates finding-specific inter-region occurrence relations into a region-level graph for region-wise prediction.
By comparison, \modelname{} defines co-occurrence at the level of individual region--finding labels and uses this structure to organize label representations.
\section{Method}
\label{sec:method}
\begin{figure*}[t]
    \centering
    \includegraphics[width=0.87\linewidth]
    {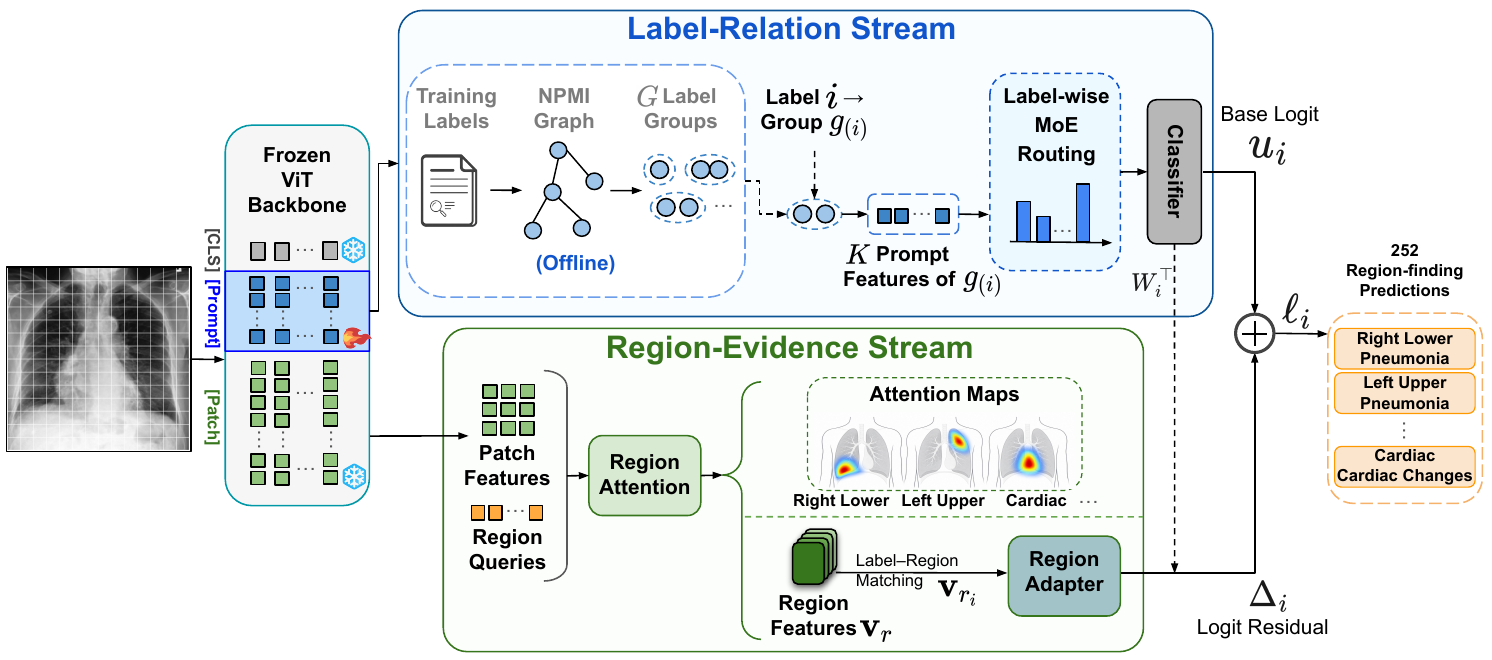}
    \caption{\footnotesize
    \textbf{Overview of \modelname{}.} }
    \label{fig:framework}
\end{figure*}

\modelname{} predicts 252 anatomical region--finding labels (e.g., left upper lung zone--pneumonia), corresponding to valid combinations of 28 findings and 22 regions defined in CI(Chest ImaGenome~\citep{wu2021imagenome}).
The framework uses two streams: a Label-Relation Stream for co-occurrence-guided label representation and a Region-Evidence Stream for region-specific visual evidence.
Both streams operate on representations from a frozen vision transformer (ViT)~\citep{dosovitskiy2021image} backbone, with visual prompt tuning (VPT)~\citep{jia2022visual} used in the Label-Relation Stream.
Fig.~\ref{fig:framework} illustrates the overall architecture.

\subsection{Label-Relation Stream}
\label{sec:label_relation}

\paragraph{Co-occurrence grouping.}
We estimate pairwise associations among region--finding labels using pointwise mutual information (PMI) computed on the training split.
Rather than relying only on how often two labels occur together, PMI accounts for the marginal prevalence of each label and quantifies how much their joint occurrence deviates from what would be expected under independence.
For labels $i$ and $j$ with $p_{ij}>0$, PMI is defined as
\begin{equation}
\mathrm{PMI}_{ij}
= \log \frac{p_{ij}}{p_i p_j},
\end{equation}
where $p_i$ and $p_j$ denote the empirical marginal probabilities of labels $i$ and $j$, respectively, and $p_{ij}$ denotes their empirical joint probability.

To compare association strengths across label pairs on a consistent scale, we use normalized pointwise mutual information (NPMI)~\citep{bouma2009}, which normalizes PMI by the self-information of the joint event
\begin{equation}
\mathrm{NPMI}_{ij}
= \frac{\mathrm{PMI}_{ij}}{-\log(p_{ij})}.
\end{equation}
NPMI is positive, zero, or negative when co-occurrence is above, equal to, or below that expected under independence, respectively. To capture positive co-occurrence structure, we retain positive NPMI values and set non-positive values to zero. Pairs never co-occurring in training ($p_{ij}=0$) have undefined PMI and NPMI and are also assigned zero affinity. The resulting affinity matrix is defined as
\begin{equation}
C_{ij} =
\begin{cases}
\max(0,\mathrm{NPMI}_{ij}), & i\neq j,\; p_{ij}>0,\\
0, & \text{otherwise}.
\end{cases}
\end{equation}
We use the NPMI-based affinity, estimated from dataset-level co-occurrence statistics, as a structural prior for representation sharing rather than as direct label-to-label interaction weights. We apply spectral clustering~\citep{spectralclustering} to embed the NPMI affinity graph into a low-dimensional spectral space reflecting its connectivity structure and partition the labels into $G$ co-occurrence groups, where $g_i\in\{1,\ldots,G\}$ denotes the fixed group assignment of label $i$.

\paragraph{Image-conditioned prompt-expert routing.}
We adapt the group-wise prompt--expert routing mechanism of ML-VPT~\citep{ma2025mlvpt} to the co-occurrence groups defined over region--finding labels.
We use deep visual prompt tuning with $K$ learnable prompt--expert pairs shared by the labels in each co-occurrence group, yielding $G\times K$ layer-specific prompts inserted into every ViT block.
The prompts are processed jointly with the CLS and patch tokens, while attention between distinct prompt tokens is masked.
We denote the resulting image-conditioned feature of the $k$-th prompt in group $g$ by $\mathbf{h}_{g,k}(x)\in\mathbb{R}^{D}$, where $D$ is the ViT feature dimension.

Each prompt feature is transformed by its corresponding expert $\mathcal{E}_{g,k}$:
\begin{equation}
\begin{aligned}
\mathbf{e}_{g,k}(x)
&= \mathcal{E}_{g,k}\!\left(\mathbf{h}_{g,k}(x)\right) \\
&= \mathbf{h}_{g,k}(x)
+ \operatorname{MLP}_{g,k}\!\left(\mathbf{h}_{g,k}(x)\right),
\end{aligned}
\label{eq:prompt_expert}
\end{equation}
where $\operatorname{MLP}_{g,k}$ is a two-layer MLP with GELU activation.
The resulting $K$ expert outputs form a set of image-conditioned candidate representations shared within each group.

For label $i$, a label-specific gate uses the $K$ prompt features in its assigned group $g_i$ to compute routing weights over the corresponding expert outputs:
\begin{equation}
\begin{aligned}
s_{i,k}(x)
&= \mathbf{w}_{i,k}^{\top}\mathbf{h}_{g_i,k}(x), \\
\pi_{i,k}(x)
&=
\frac{\exp\!\left(s_{i,k}(x)\right)}
{\sum_{t=1}^{K}\exp\!\left(s_{i,t}(x)\right)},
\end{aligned}
\label{eq:label_gate}
\end{equation}
where $s_{i,k}(x)$ is the routing score, $\mathbf{w}_{i,k}\in\mathbb{R}^{D}$ is a label-specific learnable gating vector, and $\pi_{i,k}(x)$ is the softmax-normalized routing weight over the $K$ candidates.
The label-side representation is obtained as the routing-weighted mixture of the corresponding expert outputs:
\begin{equation}
\mathbf{z}_i(x)
=
\sum_{k=1}^{K}
\pi_{i,k}(x)\mathbf{e}_{g_i,k}(x).
\label{eq:label_representation}
\end{equation}

\subsection{Region-Evidence Stream}
\label{sec:region_evidence}

To complement the label-side representations with localized visual evidence, the Region-Evidence Stream aggregates $M$ patch features $\{\mathbf p_m(x)\}_{m=1}^{M}$ from the frozen ViT using a learnable query $\mathbf{q}_r\in\mathbb{R}^{D}$ for each region $r$.
For each region, we compute region attention using the scaled cosine similarity between its query and the patch features:

\begin{equation}
\begin{aligned}
\alpha_{r,m}(x)
&=
\sqrt{D}\,
\frac{
\mathbf{q}_r^{\top}\mathbf{p}_m(x)
}{
\lVert\mathbf{q}_r\rVert_2
\lVert\mathbf{p}_m(x)\rVert_2
}, \\
\beta_{r,m}(x)
&= \sigma\!\left(\alpha_{r,m}(x)\right),
\end{aligned}
\label{eq:region_attention}
\end{equation}
where $\alpha_{r,m}(x)$ is the region-attention logit, $\beta_{r,m}(x)$ is its sigmoid weight, and $\sigma$ is the sigmoid function.
The factor $\sqrt{D}$ rescales the cosine similarities before sigmoid weighting.
Independent sigmoid weighting allows multiple patches to receive high relevance for the same region without enforcing competition among patches.

We then form the region representation by adding the region query to the attention-weighted average of the patch features:
\begin{equation}
\mathbf{v}_r(x)
=
\operatorname{LN}\!\left(
\mathbf{q}_r
+
\frac{
\sum_{m=1}^{M}\beta_{r,m}(x)\mathbf{p}_m(x)
}{
\sum_{m=1}^{M}\beta_{r,m}(x)+\epsilon
}
\right),
\label{eq:region_representation}
\end{equation}
where $\operatorname{LN}$ denotes layer normalization and $\epsilon$ is a small constant for numerical stability. The weights $\{\beta_{r,m}(x)\}_{m=1}^{M}$ provide the corresponding spatial attention map.

\subsection{Region-Matched Logit Residual}
\label{sec:matched_region_residual}

To combine the two streams at the region--finding label level, each label $i$ is matched to its anatomical region $r_i$, and its regional contribution is derived from $\mathbf{v}_{r_i}(x)$.
We transform this representation with a shared adapter $\mathcal{A}$, implemented as a two-layer bottleneck MLP with GELU, before label-specific scoring.

A linear classification head maps each label-side representation $\mathbf{z}_i(x)$ to a base logit $u_i(x)$ using the label-specific weight vector $\mathbf{w}_i$ and bias $b_i$.
We reuse $\mathbf{w}_i$ to score the adapted region
representation, yielding an additive logit residual:
\begin{equation}
\begin{aligned}
u_i(x)
&= \mathbf{w}_i^{\top}\mathbf{z}_i(x)+b_i, \\
\Delta_i(x)
&= \mathbf{w}_i^{\top}
\mathcal{A}\!\left(\mathbf{v}_{r_i}(x)\right), \\
\ell_i(x)
&= u_i(x)+\Delta_i(x).
\end{aligned}
\label{eq:region_residual}
\end{equation}
Reusing $\mathbf{w}_i$ projects the shared regional representation onto each label-specific decision direction, producing distinct residuals for labels associated with the same region.
The output projection of $\mathcal{A}$ is zero-initialized, setting $\Delta_i(x)=0$ at initialization so that the untrained region residual does not perturb the base logit.

\subsection{Training Objective}
\label{sec:training_objective}

The 252-label space is sparse and imbalanced, motivating the use of asymmetric loss (ASL)~\citep{ridnik2021asymmetric}.
We optimize the final region--finding logits as
\begin{equation}
\begin{aligned}
\mathcal{L}_{\mathrm{cls}}
&=
\operatorname{ASL}\!\left(\boldsymbol{\ell}(x),\mathbf{y}\right) \\
&=
-\sum_i
y_i(1-p_i(x))^{\gamma^{+}}\log p_i(x) \\
&\quad
-\sum_i
(1-y_i)p_i(x)^{\gamma^{-}}
\log\!\left(1-p_i(x)\right),
\end{aligned}
\label{eq:classification_loss}
\end{equation}
where $\mathbf{y}\in\{0,1\}^{252}$ is the ground-truth region--finding label vector, $p_i(x)=\sigma(\ell_i(x))$, and $\gamma^{+}$ and $\gamma^{-}$ control the focusing of the positive and negative terms, respectively.

To make the routing scores reflect label presence in addition to relative expert preference, we apply an auxiliary gating objective only during training, while final predictions are computed from $\ell_i(x)$ in Eq. \ref{eq:region_residual}.
For each label, we form an auxiliary label logit as a routing-weighted combination of its $K$ routing scores and apply ASL:
\begin{equation}
\begin{aligned}
s_i^{\mathrm{gate}}(x)
&=
\sum_{k=1}^{K}\pi_{i,k}(x)s_{i,k}(x), \\
\mathcal{L}_{\mathrm{gate}}
&=
\operatorname{ASL}\!\left(\mathbf{s}^{\mathrm{gate}}(x),\mathbf{y}\right).
\end{aligned}
\label{eq:gate_loss}
\end{equation}

We further supervise the Region-Evidence Stream at both patch and region levels. At the patch level, projected anatomical boxes provide binary masks for binary cross-entropy (BCE) with logits supervision of the attention logits $\alpha_{r,m}(x)$ via $\mathcal{L}_{\mathrm{attn}}$. At the region level, a shared linear head predicts whether any target finding is present in each region, with a positive target assigned if any associated region--finding label is positive, and is supervised with BCE via $\mathcal{L}_{\mathrm{reg}}$.

The complete objective is
\begin{equation}
\begin{aligned}
\mathcal{L}_{\mathrm{total}}
&=
\mathcal{L}_{\mathrm{cls}}
+ \lambda_{\mathrm{gate}}\mathcal{L}_{\mathrm{gate}}
\\
&\quad
+ \lambda_{\mathrm{attn}}\mathcal{L}_{\mathrm{attn}}
+ \lambda_{\mathrm{reg}}\mathcal{L}_{\mathrm{reg}}.
\end{aligned}
\label{eq:total_loss}
\end{equation}

\section{Experiments}
\label{sec:exp}
\subsection{Experimental Setup}

\paragraph{Dataset and label space.}
We use 241,451 frontal chest radiographs from 63,764 patients in MIMIC-CXR~\citep{mimic-cxr}, annotated with regions and findings from CI~\citep{wu2021imagenome}. The official split contains 236,089/1,959/3,403 (train/val/test) images. Retaining region--finding pairs with at least 80 positive training examples yields 252 targets spanning 28 findings and 22 anatomical regions (Appendix~\ref{apd:first}). For localization, we use bounding-box annotations from the MIMIC-CXR test split, MS-CXR~\citep{ms-cxr}, and the CI gold standard; MS-CXR and the gold standard are excluded from training at the image level. External finding-level evaluation uses the official VinDr-CXR~\citep{nguyen2022vindr} and CheXpert~\citep{irvin2019chexpert} test sets without target-domain adaptation. To handle nested anatomy, broader-region targets are masked during training when a more specific positive descendant is available, and positive predictions are propagated to anatomical ancestors at evaluation. The same hierarchy handling is applied to all methods.

\paragraph{Baselines.}
For the 252-label classification task, we evaluate linear probing (LP) and full fine-tuning (FFT) with ImageNet-pretrained ViT-B/16~\citep{dosovitskiy2021image} and the CXR foundation encoders RAD-DINO-MAIRA-2 (RAD-DINO-2)~\citep{bannur2024maira} and EVA-X~\citep{yao2025evax}. With EVA-X as the common backbone, we further compare VPT-Shallow, VPT-Deep~\citep{jia2022visual}, and ML-VPT~\citep{ma2025mlvpt} for label-relation prompt modeling. Localization baselines are ChEX~\citep{mueller2024chex} and MedGrounder~\citep{medgrounder2025}; finding-level evaluation on the CI gold standard and external datasets additionally includes the zero-shot CXR baselines CheXzero~\citep{tiu2022chexzero}, RadZero~\citep{park2025radzero}, and CARZero~\citep{lai2024carzero}.

\paragraph{Implementation details.}
\modelname{} uses a frozen pretrained EVA-X backbone with $G=8$ co-occurrence groups, $K=5$ prompt--expert pairs per group, and 22 learned region queries; sensitivity to $G$ and $K$ is reported in Appendix~\ref{apd:first}. The Region-Evidence Stream uses all $M=196$ patch tokens on the fixed $14\times14$ grid.
We train for 20 epochs with AdamW, a batch size of 512, a learning rate of $5\times10^{-4}$, weight decay of $10^{-2}$, and a OneCycle schedule.
For ASL~\citep{liu2021query2label,li2023patchct,ma2025mlvpt}, we use $\gamma^{+}=0$ and $\gamma^{-}=2$, with $\lambda_{\mathrm{gate}}=1.0$, $\lambda_{\mathrm{attn}}=0.5$, and $\lambda_{\mathrm{reg}}=0.3$.
Baseline-specific settings are provided in Appendix~\ref{apd:second}.

\paragraph{Evaluation protocol.}
We report macro AUROC and AUPRC as primary classification metrics, along with macro F1, micro F1, and trainable parameter count; F1 scores use a common fixed threshold of 0.5 across all methods.
For all methods in Table~\ref{tab:main_results}, 95\% confidence intervals for AUROC and AUPRC are estimated from 10,000 paired patient-level bootstrap resamples. Spatial alignment is evaluated using pointing-game hit rate (P.~Hit)~\citep{zhang2016topdown}.
For classification models, target-specific patch saliency is computed using gradient--activation attribution~\citep{ancona2018towards}, with a hit defined when the highest-saliency patch center falls within the ground-truth bounding box.
Further details are provided in Appendix~\ref{apd:third}.

\subsection{Overall Comparison}

\begin{table*}[t]
\centering
\caption{
Classification and spatial attribution on the 252 region--finding labels in the MIMIC-CXR test split.
P.~Hit measures target-specific attribution against CI (Chest ImaGenome) anatomical annotations.
}
\label{tab:main_results}

\scriptsize
\renewcommand{\arraystretch}{1.12}

\resizebox{\textwidth}{!}{
\begin{tabular}{@{}llcccccc@{}}
\toprule
\textbf{Method} &
\textbf{Backbone} &
\textbf{Params.} &
\textbf{AUROC} &
\textbf{AUPRC} &
\textbf{Macro F1} &
\textbf{Micro F1} &
\textbf{P. Hit} \\
\midrule

\multirow{3}{*}{LP}
& ViT-B/16
& 0.19M
& \makecell{0.6574 \\ {\scriptsize(.6389--.6695)}}
& \makecell{0.0907 \\ {\scriptsize(.0895--.1079)}}
& 0.0424
& 0.3961
& 0.1810 \\

& RAD-DINO-2
& 0.19M
& \makecell{0.7730 \\ {\scriptsize(.7557--.7848)}}
& \makecell{0.1430 \\ {\scriptsize(.1408--.1635)}}
& 0.0690
& 0.4682
& 0.2948 \\

& EVA-X
& 0.19M
& \makecell{0.7872 \\ {\scriptsize(.7701--.7987)}}
& \makecell{0.1531 \\ {\scriptsize(.1507--.1742)}}
& 0.0762
& 0.4724
& 0.3768 \\

\midrule

\multirow{3}{*}{FFT}
& ViT-B/16
& 85.99M
& \makecell{0.7526 \\ {\scriptsize(.7348--.7650)}}
& \makecell{0.1313 \\ {\scriptsize(.1294--.1508)}}
& 0.0548
& 0.4661
& 0.2627 \\

& RAD-DINO-2
& 86.77M
& \makecell{0.8035 \\ {\scriptsize(.7867--.8144)}}
& \makecell{0.1624 \\ {\scriptsize(.1598--.1840)}}
& 0.0778
& 0.5042
& 0.3425 \\

& EVA-X
& 86.04M
& \makecell{0.8075 \\ {\scriptsize(.7906--.8180)}}
& \makecell{0.1639 \\ {\scriptsize(.1612--.1857)}}
& 0.0883
& 0.5196
& 0.3872 \\

\midrule

VPT-Shallow
& EVA-X
& 0.22M
& \makecell{0.7816 \\ {\scriptsize(.7640--.7951)}}
& \makecell{0.1370 \\ {\scriptsize(.1352--.1571)}}
& 0.0732
& 0.5282
& 0.3779 \\

VPT-Deep
& EVA-X
& 0.56M
& \makecell{0.7947 \\ {\scriptsize(.7776--.8062)}}
& \makecell{0.1418 \\ {\scriptsize(.1392--.1637)}}
& 0.0783
& 0.5351
& 0.4047 \\

ML-VPT
& EVA-X
& 2.08M
& \makecell{0.8283 \\ {\scriptsize(.8113--.8384)}}
& \makecell{0.1831 \\ {\scriptsize(.1801--.2058)}}
& 0.1398
& 0.5536
& 0.5483 \\
\cmidrule(lr){1-8}

\textbf{CoRE}
& EVA-X
& 2.09M
& \makecell{\textbf{0.8430} \\ {\scriptsize\textbf{(.8261--.8531)}}}
& \makecell{\textbf{0.2024} \\ {\scriptsize\textbf{(.1961--.2295)}}}
& \textbf{0.1546}
& \textbf{0.5675}
& \textbf{0.5671} \\

\bottomrule
\end{tabular}
}
\end{table*}

As shown in Table~\ref{tab:main_results}, \modelname{} achieves the best performance across all classification metrics under the same 252-label prediction setting.
With EVA-X as the common backbone, \modelname{} substantially outperforms full fine-tuning in AUROC (0.8075 to 0.8430) and AUPRC (0.1639 to 0.2024), despite training only 2.09M parameters (97.6\% fewer trainable parameters).
\modelname{} also consistently improves over standard prompt tuning (VPT-Shallow and VPT-Deep~\citep{jia2022visual}), indicating that its gains extend beyond parameter-efficient adaptation alone. The most direct comparison is ML-VPT~\citep{ma2025mlvpt}, which explicitly incorporates label relationships into prompt-based adaptation.
Under a nearly identical trainable parameter budget (2.09M vs.\ 2.08M), \modelname{} improves AUROC from 0.8283 to 0.8430 and AUPRC from 0.1831 to 0.2024. The corresponding paired differences are 0.0147 (95\% CI [0.0083, 0.0209]) and 0.0193 (95\% CI [0.0077, 0.0263]), respectively. \modelname{} also improves macro F1 from 0.1398 to 0.1546 and micro F1 from 0.5536 to 0.5675.
\modelname{} further improves P.~Hit from 0.5483 to 0.5671, showing that the classification gains are accompanied by better spatial correspondence of target-specific evidence.
These results demonstrate consistent gains over existing relation-aware prompt modeling under a matched parameter budget.

\subsection{Ablation Studies}
We ablate the two-stream architecture, label grouping, and auxiliary objectives to assess the contribution of each design component.

\begin{table}[htbp]
\centering
\caption{
Ablation of the two-stream architecture and label grouping on the MIMIC-CXR test split. L-S and R-S denote the Label-Relation and Region-Evidence Streams.
}
\label{tab:ablation}
\scriptsize
\setlength{\tabcolsep}{4pt}
\renewcommand{\arraystretch}{1.15}

\begin{adjustbox}{max width=\columnwidth}
\begin{tabular}{@{}lcccc@{}}
\toprule
\textbf{Configuration} &
\textbf{L-S} &
\textbf{R-S} &
\textbf{AUROC} &
\textbf{AUPRC} \\
\midrule

w/o L-S
& \xmark
& \cmark
& \makecell{0.8250 \\ {\scriptsize(.8081--.8351)}}
& \makecell{0.1800 \\ {\scriptsize(.1722--.2048)}} \\

w/o R-S
& \cmark
& \xmark
& \makecell{0.8408 \\ {\scriptsize(.8241--.8514)}}
& \makecell{0.2003 \\ {\scriptsize(.1925--.2251)}} \\

\midrule

Single ($G=1$)
& \cmark
& \cmark
& \makecell{0.8335 \\ {\scriptsize(.8167--.8438)}}
& \makecell{0.1902 \\ {\scriptsize(.1840--.2175)}} \\

Random ($G=8$)
& \cmark
& \cmark
& \makecell{0.8423 \\ {\scriptsize(.8255--.8526)}}
& \makecell{0.1993 \\ {\scriptsize(.1932--.2268)}} \\

Conditional ($G=8$)
& \cmark
& \cmark
& \makecell{0.8411 \\ {\scriptsize(.8244--.8517)}}
& \makecell{0.1986 \\ {\scriptsize(.1926--.2261)}} \\

\textbf{NPMI ($G=8$, CoRE)}
& \cmark
& \cmark
& \makecell{\textbf{0.8430} \\ {\scriptsize\textbf{(.8261--.8531)}}}
& \makecell{\textbf{0.2024} \\ {\scriptsize\textbf{(.1961--.2295)}}} \\

\bottomrule
\end{tabular}
\end{adjustbox}
\end{table}

\paragraph{Two-stream architecture and grouping.}
Table~\ref{tab:ablation} evaluates the two streams and grouping strategies.
Removing either stream reduces performance, with a larger drop without the \textit{Label-Relation Stream}, while the full model performs best, supporting the complementary roles of label-side representation learning and region-matched visual evidence.

To isolate the effect of label grouping, the single-group baseline ($G=1$) retains label-specific prompt--expert routing ($K=5$) but shares the same candidates across all labels.
All multi-group variants ($G=8$) outperform this baseline, indicating a benefit from group-wise representation sharing.
We compare Random, Conditional grouping adapted from ML-VPT~\citep{ma2025mlvpt}, and our NPMI-based grouping; Conditional uses conditional co-occurrence affinities followed by spectral clustering.
While the three strategies perform similarly, NPMI yields the highest point estimates in Table~\ref{tab:ablation}; results over three training seeds are reported in Appendix \ref{apd:first}.

\paragraph{Auxiliary objectives.}
Table~\ref{tab:loss_ablation} evaluates supervision for prompt routing and anatomical evidence. The full objective achieves the highest AUROC and AUPRC, with both decreasing when any individual auxiliary loss is removed. Adding anatomical supervision to $\mathcal{L}_{\mathrm{gate}}$ also improves P.~Hit, and removing $\mathcal{L}_{\mathrm{attn}}$ produces the largest P.~Hit reduction among the leave-one-out variants, consistent with its direct supervision of spatial attention. Although removing $\mathcal{L}_{\mathrm{gate}}$ slightly increases P.~Hit, it reduces both AUROC and AUPRC, indicating different contributions of routing and anatomical supervision to classification and spatial alignment. Overall, the full objective provides the strongest classification performance while maintaining competitive spatial alignment.

\begin{table}[htbp]
\centering
\caption{Ablation of the auxiliary training objectives in the CoRE architecture on the MIMIC-CXR test split.}
\label{tab:loss_ablation}
\scriptsize
\renewcommand{\arraystretch}{1.12}
\begin{tabular}{@{}cccccc@{}}
\toprule
$\mathcal{L}_{\mathrm{gate}}$ &
$\mathcal{L}_{\mathrm{attn}}$ &
$\mathcal{L}_{\mathrm{reg}}$ &
\textbf{AUROC} &
\textbf{AUPRC} &
\textbf{P. Hit} \\
\midrule
\xmark & \xmark & \xmark & 0.8378 & 0.1917 & 0.5479 \\
\cmark & \xmark & \xmark & 0.8409 & 0.1941 & 0.5493 \\
\cmark & \cmark & \xmark & 0.8414 & 0.1943 & 0.5684 \\
\cmark & \xmark & \cmark & 0.8416 & 0.1989 & 0.5523 \\
\xmark & \cmark & \cmark & 0.8390 & 0.1946 & \textbf{0.5745} \\
\cmark & \cmark & \cmark & \textbf{0.8430} & \textbf{0.2024} & 0.5671 \\
\bottomrule
\end{tabular}
\end{table}

\begin{table}[htbp]
\centering
\caption{
Finding localization performance on MS-CXR and the CI gold standard measured by P.~Hit. Image and Text indicate the input modalities used by each method.}
\label{tab:localization}
\scriptsize
\renewcommand{\arraystretch}{1.12}

\begin{tabular}{@{}lcccc@{}}
\toprule
\multirow{2}{*}{\textbf{Method}} &
\multirow{2}{*}{\textbf{Image}} &
\multirow{2}{*}{\textbf{Text}} &
\multicolumn{2}{c}{\textbf{P. Hit}} \\
\cmidrule(l){4-5}
& & & \textbf{MS-CXR} & \textbf{CI} \\
\midrule

ChEX & \cmark & \cmark & 0.7724 & 0.8381 \\
MedGrounder & \cmark & \cmark & 0.7603 & 0.7895 \\
ML-VPT & \cmark & \xmark & 0.7595 & 0.8702 \\
CoRE w/o R-S & \cmark & \xmark & 0.7655 & 0.8861 \\

\midrule

\textbf{CoRE} & \cmark & \xmark & \textbf{0.7853} & \textbf{0.9045} \\

\bottomrule
\end{tabular}
\end{table}

\subsection{Finding-Specific Spatial Alignment}

We evaluate whether region--finding prediction leads to spatially aligned evidence for individual findings using bounding-box annotations from MS-CXR and the CI gold standard. We include phrase-grounding models as reference comparisons using P.~Hit, which provides a common point-localization measure for both saliency and box outputs. For classification models, finding-level saliency is derived from the maximum logit across region--finding outputs for each finding. For phrase-grounding models, we use the annotated phrase on MS-CXR and the corresponding region--finding phrase on CI, with P.~Hit computed from the center of the top-ranked predicted box~\citep{liu2021relation,medgrounder2025}. As shown in Table~\ref{tab:localization}, \modelname{} achieves the highest P.~Hit on both MS-CXR (0.7853) and CI (0.9045). Adding the \textit{Region-Evidence Stream} improves P.~Hit from 0.7655 to 0.7853 on MS-CXR and from 0.8861 to 0.9045 on CI, demonstrating that region-matched visual evidence improves spatial alignment beyond the \textit{Label-Relation Stream} alone.

\begin{figure*}[t]
    \centering
    \includegraphics[width=0.76\textwidth]{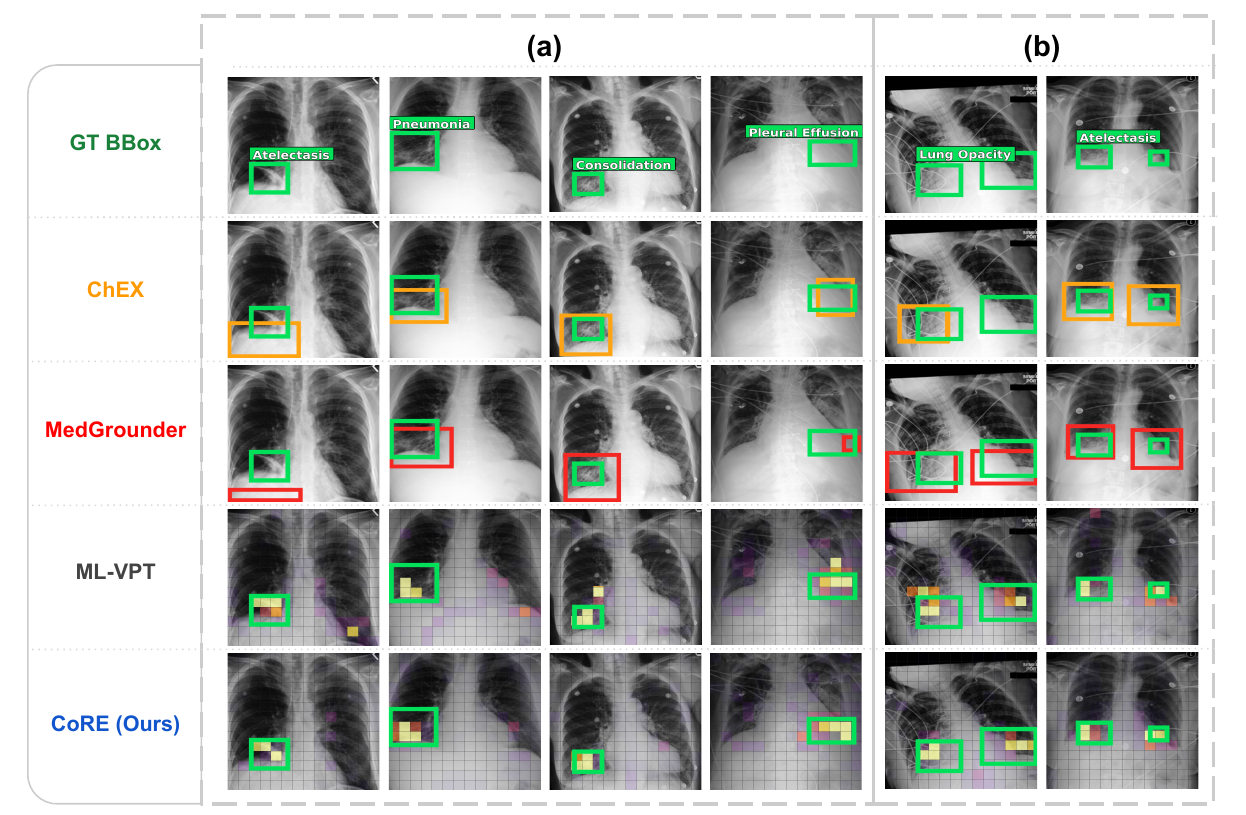}
\caption{\footnotesize
\textbf{Qualitative comparison of finding localization on MS-CXR.}
Columns show representative examples of \textbf{(a)} single findings with a single box and \textbf{(b)} single findings with multiple boxes.
Rows show ground-truth boxes, predicted boxes from ChEX~\citep{mueller2024chex} and MedGrounder~\citep{medgrounder2025}, and patch-saliency maps from ML-VPT~\citep{ma2025mlvpt} and \modelname{}.
}
    \label{fig:visual}
\end{figure*}

\begin{table}[htbp]
\centering
\caption{
Finding-level performance on the internal CI evaluation and external VinDr-CXR and CheXpert datasets.
}
\label{tab:finding_level}
\scriptsize
\setlength{\tabcolsep}{5.0pt}
\renewcommand{\arraystretch}{1.12}

\begin{tabular}{@{}lccc@{}}
\toprule
\textbf{Method} &
\makecell{\textbf{CI} \\ \scriptsize(28 labels)} &
\makecell{\textbf{VinDr-CXR} \\ \scriptsize(9 labels)} &
\makecell{\textbf{CheXpert} \\ \scriptsize(8 labels)} \\
\midrule

\multicolumn{4}{c}{\textbf{AUROC}} \\
\cmidrule(lr){1-4}

CheXzero & 0.6731 & 0.8444 & 0.8428 \\
RadZero  & 0.7877 & 0.8859 & 0.8968 \\
CARZero  & 0.7820 & 0.9202 & 0.9100 \\

\hdashline

VPT-Deep & 0.8010 & 0.8771 & 0.8667 \\
ML-VPT   & 0.8319 & 0.9140 & 0.8983 \\
\textbf{CoRE} & \textbf{0.8442} & \textbf{0.9313} & \textbf{0.9125} \\

\midrule

\multicolumn{4}{c}{\textbf{AUPRC}} \\
\cmidrule(lr){1-4}

CheXzero & 0.2130 & 0.3496 & 0.5465 \\
RadZero  & 0.2885 & 0.4372 & 0.5810 \\
CARZero  & 0.2948 & 0.4708 & 0.6521 \\

\hdashline

VPT-Deep & 0.3098 & 0.4182 & 0.5712 \\
ML-VPT   & 0.3564 & 0.4777 & 0.6540 \\
\textbf{CoRE} & \textbf{0.3748} & \textbf{0.4978} & \textbf{0.6653} \\

\bottomrule
\end{tabular}
\end{table}
\subsection{Finding-Level Transfer and Generalization}

We further evaluate whether fine-grained region--finding predictions remain effective when aggregated into conventional finding-level predictions, including on datasets without anatomically specified labels. For each finding, we take the maximum logit across its anatomical regions to obtain a finding-level prediction. We evaluate all 28 findings on the CI gold standard, 9 aligned findings on VinDr-CXR, and 8 on CheXpert.\footnote{The aligned findings are atelectasis, cardiomegaly, consolidation, interstitial lung disease, lung opacity, nodule/mass, pleural effusion, pneumothorax, and pneumonia for VinDr-CXR; and cardiomegaly, atelectasis, lung opacity, edema, consolidation, pneumonia, pneumothorax, and pleural effusion for CheXpert.} As shown in Table~\ref{tab:finding_level}, \modelname{} achieves the highest AUROC and AUPRC across all three datasets. Compared with ML-VPT, \modelname{} improves AUROC/AUPRC from 0.8319/0.3564 to 0.8442/0.3748 on CI, 0.9140/0.4777 to 0.9313/0.4978 on VinDr-CXR, and 0.8983/0.6540 to 0.9125/0.6653 on CheXpert. \modelname{} also achieves higher performance than the evaluated zero-shot CXR baselines on both external datasets, showing that its fine-grained predictions remain effective after finding-level aggregation and under external domain shift.

\subsection{Visualization of Finding-Specific Evidence}

Figure~\ref{fig:visual} visualizes finding-specific evidence in single- and multi-box localization settings. For ChEX and MedGrounder, we use the annotated phrase as the query and show the predicted boxes; for ML-VPT and \modelname{}, we visualize gradient--activation saliency from the region--finding output corresponding to the annotated target. For single-box findings in (a), the grounding models generally identify the target region but can produce spatially shifted or oversized boxes. In the right lower lung consolidation case, both grounding models extend beyond the ground-truth box, whereas ML-VPT and \modelname{} place strong saliency within the annotated region. \modelname{} shows a more concentrated response, while ML-VPT exhibits an isolated off-target activation. For multi-box findings in (b), grounding predictions may only partially cover the annotated regions or extend beyond their boundaries. Both image-only models respond to relevant regions, but \modelname{} shows more concentrated saliency within the annotated areas with less spillover. These target-conditioned examples qualitatively complement the finding-level spatial alignment results in Table~\ref{tab:localization}; representative error cases are shown in Appendix~\ref{fourth}.
\section{Conclusion}
\label{sec:conclusion}
We introduced \modelname{}, a framework for fine-grained chest X-ray prediction at the anatomical region--finding level. \modelname{} combines group-wise label representation learning with region-specific visual evidence, improving region--finding classification while using only a small fraction of trainable parameters compared with full fine-tuning. Its region-evidence mechanism also improves the spatial alignment of finding-specific evidence with annotated abnormalities. When region--finding outputs are aggregated to conventional finding-level predictions, \modelname{} remains effective on both internal and external datasets without target-domain adaptation. While the choice of grouping strategy has a relatively small effect on predictive performance, NPMI yields a data-driven partition of anatomically specified labels into relatively balanced groups that reflect the observed co-occurrence structure. Together, these results highlight the benefit of combining structured label sharing with anatomically grounded visual evidence for fine-grained CXR analysis.

Several limitations remain. \modelname{} currently relies on static co-occurrence statistics to organize label sharing, and fine-grained region--finding supervision and validation are largely limited to a single anatomically annotated dataset. In addition, spatial alignment is evaluated using bounding-box-based criteria rather than dense localization. Future work could explore adaptive region--finding interactions, broader cross-dataset validation with anatomical annotations, and more precise spatial supervision.

\acks{This work was supported by the TIPS (Tech Incubator Program for Startup) Program of the Ministry of SMEs and Startups and Korea Startup Foundation under Grant No. RS-2025-25467010.}
\clearpage

\bibliography{ref}
\clearpage
\appendix

\begin{table*}[!b]
\centering
\caption{
All 252 region--finding labels are shown in a finding--by--region incidence matrix; filled cells indicate retained pairs.
Region abbreviations:
LUL/LML/LLL/LAZ/LCP/LHil/LDia (left upper/mid/lower lung zone, apical zone, costophrenic angle, hilar, hemidiaphragm);
RUL/RML/RLL/RAZ/RCP/RHil/RDia (right counterparts);
LLu/RLu (left/right lung);
Car (cardiac);
Med/UMed (mediastinum/upper mediastinum);
AoA (aortic arch);
Tr (trachea);
Sp (spine).
}
\label{tab:label_incidence}
\scriptsize
\setlength{\tabcolsep}{2.2pt}
\renewcommand{\arraystretch}{1.05}
\resizebox{\textwidth}{!}{%
\begin{tabular}{@{}l*{22}{c}c@{}}
\toprule
\textbf{Finding} &
LUL & LML & LLL & LAZ & LCP & LHil & LDia &
RUL & RML & RLL & RAZ & RCP & RHil & RDia &
LLu & RLu & Car & Med & UMed & AoA & Tr & Sp &
\# \\
\midrule
alveolar hemorrhage & $\bullet$ & $\bullet$ & $\bullet$ &  &  &  &  & $\bullet$ & $\bullet$ & $\bullet$ &  &  &  &  & $\bullet$ & $\bullet$ &  &  &  &  &  &  & 8 \\
aspiration & $\bullet$ & $\bullet$ & $\bullet$ &  &  & $\bullet$ &  & $\bullet$ & $\bullet$ & $\bullet$ &  &  & $\bullet$ &  & $\bullet$ & $\bullet$ &  &  &  &  &  &  & 10 \\
copd/emphysema & $\bullet$ & $\bullet$ & $\bullet$ &  &  & $\bullet$ & $\bullet$ & $\bullet$ & $\bullet$ & $\bullet$ &  &  & $\bullet$ & $\bullet$ & $\bullet$ & $\bullet$ &  &  &  &  &  &  & 12 \\
fluid overload/heart failure & $\bullet$ & $\bullet$ & $\bullet$ &  & $\bullet$ & $\bullet$ &  & $\bullet$ & $\bullet$ & $\bullet$ &  & $\bullet$ & $\bullet$ &  & $\bullet$ & $\bullet$ & $\bullet$ &  &  &  &  &  & 13 \\
goiter &  &  &  &  &  &  &  &  &  &  &  &  &  &  &  &  &  &  & $\bullet$ &  & $\bullet$ &  & 2 \\
granulomatous disease & $\bullet$ & $\bullet$ & $\bullet$ & $\bullet$ &  & $\bullet$ &  & $\bullet$ & $\bullet$ & $\bullet$ & $\bullet$ &  & $\bullet$ &  & $\bullet$ & $\bullet$ &  &  &  &  &  &  & 12 \\
interstitial lung disease & $\bullet$ & $\bullet$ & $\bullet$ &  &  &  &  & $\bullet$ & $\bullet$ & $\bullet$ &  &  &  &  & $\bullet$ & $\bullet$ &  &  &  &  &  &  & 8 \\
lung cancer & $\bullet$ & $\bullet$ & $\bullet$ &  &  & $\bullet$ &  & $\bullet$ & $\bullet$ & $\bullet$ &  &  & $\bullet$ &  & $\bullet$ & $\bullet$ &  & $\bullet$ &  &  &  &  & 11 \\
pericardial effusion &  &  &  &  &  &  &  &  &  &  &  &  &  &  &  &  & $\bullet$ &  &  &  &  &  & 1 \\
pneumonia & $\bullet$ & $\bullet$ & $\bullet$ & $\bullet$ &  & $\bullet$ &  & $\bullet$ & $\bullet$ & $\bullet$ & $\bullet$ &  & $\bullet$ &  & $\bullet$ & $\bullet$ &  &  &  &  &  &  & 12 \\
aorta changes &  &  &  &  &  &  &  &  &  &  &  &  &  &  &  &  &  & $\bullet$ & $\bullet$ & $\bullet$ &  &  & 3 \\
atelectasis & $\bullet$ & $\bullet$ & $\bullet$ &  & $\bullet$ & $\bullet$ &  & $\bullet$ & $\bullet$ & $\bullet$ &  & $\bullet$ & $\bullet$ &  & $\bullet$ & $\bullet$ &  &  &  &  &  &  & 12 \\
bone changes &  &  &  &  &  &  &  &  &  &  &  &  &  &  &  &  &  &  &  &  &  & $\bullet$ & 1 \\
bronchiectasis &  & $\bullet$ & $\bullet$ &  &  &  &  & $\bullet$ & $\bullet$ & $\bullet$ &  &  &  &  & $\bullet$ & $\bullet$ &  &  &  &  &  &  & 7 \\
cardiac changes &  &  &  &  &  &  &  &  &  &  &  &  &  &  &  &  & $\bullet$ &  &  &  &  &  & 1 \\
consolidation & $\bullet$ & $\bullet$ & $\bullet$ &  & $\bullet$ & $\bullet$ &  & $\bullet$ & $\bullet$ & $\bullet$ &  & $\bullet$ & $\bullet$ &  & $\bullet$ & $\bullet$ &  &  &  &  &  &  & 12 \\
cyst/bullae &  &  & $\bullet$ & $\bullet$ &  &  &  & $\bullet$ & $\bullet$ & $\bullet$ & $\bullet$ &  &  &  & $\bullet$ & $\bullet$ &  &  &  &  &  &  & 8 \\
hernia &  &  &  &  &  &  & $\bullet$ &  &  &  &  &  &  & $\bullet$ &  &  & $\bullet$ & $\bullet$ &  &  &  &  & 4 \\
hyperaeration & $\bullet$ & $\bullet$ & $\bullet$ &  &  &  & $\bullet$ & $\bullet$ & $\bullet$ & $\bullet$ &  &  &  & $\bullet$ & $\bullet$ & $\bullet$ &  &  &  &  &  &  & 10 \\
ild pattern &  & $\bullet$ & $\bullet$ &  &  &  &  &  & $\bullet$ & $\bullet$ &  &  &  &  & $\bullet$ & $\bullet$ &  &  &  &  &  &  & 6 \\
lung opacity & $\bullet$ & $\bullet$ & $\bullet$ & $\bullet$ & $\bullet$ & $\bullet$ &  & $\bullet$ & $\bullet$ & $\bullet$ & $\bullet$ & $\bullet$ & $\bullet$ &  & $\bullet$ & $\bullet$ &  &  &  &  &  &  & 14 \\
mass/nodule & $\bullet$ & $\bullet$ & $\bullet$ & $\bullet$ & $\bullet$ & $\bullet$ &  & $\bullet$ & $\bullet$ & $\bullet$ & $\bullet$ & $\bullet$ & $\bullet$ &  & $\bullet$ & $\bullet$ &  &  & $\bullet$ &  &  &  & 15 \\
mediastinal changes &  &  &  &  &  & $\bullet$ &  &  &  &  &  &  & $\bullet$ &  &  &  & $\bullet$ & $\bullet$ & $\bullet$ & $\bullet$ & $\bullet$ &  & 7 \\
pleural effusion & $\bullet$ & $\bullet$ & $\bullet$ & $\bullet$ & $\bullet$ &  &  & $\bullet$ & $\bullet$ & $\bullet$ & $\bullet$ & $\bullet$ &  &  & $\bullet$ & $\bullet$ &  &  &  &  &  &  & 12 \\
pleural scarring & $\bullet$ & $\bullet$ & $\bullet$ & $\bullet$ & $\bullet$ & $\bullet$ &  & $\bullet$ & $\bullet$ & $\bullet$ & $\bullet$ & $\bullet$ & $\bullet$ &  & $\bullet$ & $\bullet$ &  &  &  &  &  &  & 14 \\
pneumothorax &  & $\bullet$ & $\bullet$ & $\bullet$ & $\bullet$ &  &  & $\bullet$ & $\bullet$ & $\bullet$ & $\bullet$ & $\bullet$ &  &  & $\bullet$ & $\bullet$ &  &  &  &  &  &  & 11 \\
pulmonary edema & $\bullet$ & $\bullet$ & $\bullet$ &  & $\bullet$ & $\bullet$ &  & $\bullet$ & $\bullet$ & $\bullet$ &  & $\bullet$ & $\bullet$ &  & $\bullet$ & $\bullet$ &  &  &  &  &  &  & 12 \\
vascular changes & $\bullet$ & $\bullet$ & $\bullet$ &  &  & $\bullet$ &  & $\bullet$ & $\bullet$ & $\bullet$ &  &  & $\bullet$ &  & $\bullet$ & $\bullet$ & $\bullet$ & $\bullet$ & $\bullet$ & $\bullet$ &  &  & 14 \\
\bottomrule
\end{tabular}%
}
\end{table*}
\section{Label Space and Grouping Analysis}\label{apd:first}
\subsection{Region--finding label space}
We follow the 28-finding vocabulary used in prior work with CI annotations~\citep{moon2026clinically}. Among combinations of these findings and CI anatomical regions, we retain pairs with at least 80 positive examples in the training split, yielding 252 labels spanning 22 regions. Accordingly, the label space is restricted to sufficiently supported pairs within this vocabulary.

Table~\ref{tab:label_incidence} enumerates the full label set as a finding--by--region incidence matrix, where each marked cell corresponds to one retained region--finding label. When a more specific region--finding label is positive, the corresponding broader-region target is masked during training. At evaluation, each broader-region score is replaced by the maximum of its own score and those of its same-finding anatomical descendants on the same side, with the same upward closure applied to the ground-truth labels.

\subsection{Grouping balance and composition}
We compare the group-size distributions and compositions of Conditional and NPMI grouping.
Following prior conditional co-occurrence grouping methods~\citep{xu2022boosting,ma2025mlvpt}, Conditional grouping constructs pairwise associations as $S_{ij}=P(y_j=1\mid y_i=1)$, applies element-wise root smoothing, symmetrizes the affinity as $(S^{1/\tau}+(S^{1/\tau})^{\top})/2$, and partitions the labels using spectral clustering.

We compare group-size distributions produced by Conditional and NPMI grouping across $G\in\{4,6,8,10\}$. As shown in Table~\ref{tab:grouping_strategy}, NPMI produces a narrower group-size range than Conditional grouping for all $G$ values. At the selected setting of $G=8$, group sizes range from 16 to 51 labels for NPMI, compared with 11 to 76 labels for Conditional grouping. Table~\ref{tab:group_composition} further compares the finding composition of the two grouping strategies at $G=8$.

\begin{table*}[t]
\centering
\caption{
Composition of Conditional and NPMI co-occurrence groups ($G{=}8$).
\#labels is the number of region--finding labels in each group.
Findings lists unique finding names after removing the anatomical region; the same finding may appear via multiple regions within a group.
}
\label{tab:group_composition}
\small 
\setlength{\tabcolsep}{4pt} 
\renewcommand{\arraystretch}{1.1} 

\begin{tabular}{@{}ccc>{\raggedright\arraybackslash}p{0.7\textwidth}@{}}
\toprule
\textbf{Group ID} & \textbf{\#labels} & \textbf{\#find.} & \textbf{Findings} \\
\midrule

\multicolumn{4}{@{}l}{\textbf{Conditional}} \\
\addlinespace[2pt]

$g$1 & 30 & 5 & granulomatous disease, mass/nodule, lung opacity, pleural scarring, pneumonia \\
$g$2 & 30 & 3 & copd/emphysema, hyperaeration, cyst/bullae \\
$g$3 & 31 & 4 & interstitial lung disease, pleural scarring, bronchiectasis, ild pattern \\
$g$4 & 15 & 3 & pneumothorax, pleural effusion, mediastinal changes \\
$g$5 & 76 & 10 & atelectasis, fluid overload/heart failure, lung opacity, pleural effusion, pulmonary edema, vascular changes, consolidation, mass/nodule, cardiac changes, pericardial effusion \\
$g$6 & 11 & 1 & lung cancer \\
$g$7 & 38 & 5 & alveolar hemorrhage, aspiration, consolidation, pneumonia, pulmonary edema \\
$g$8 & 21 & 7 & pleural scarring, hernia, mediastinal changes, vascular changes, aorta changes, goiter, bone changes \\

\addlinespace[3pt]
\midrule
\addlinespace[2pt]

\multicolumn{4}{@{}l}{\textbf{NPMI}} \\
\addlinespace[2pt]

$g$1 & 33 & 4 & copd/emphysema, hyperaeration, cyst/bullae, pleural scarring \\
$g$2 & 42 & 5 & mass/nodule, granulomatous disease, lung cancer, lung opacity, pneumonia \\
$g$3 & 16 & 3 & pneumothorax, mediastinal changes, pleural effusion \\
$g$4 & 23 & 4 & pleural effusion, atelectasis, lung opacity, consolidation \\
$g$5 & 37 & 7 & fluid overload/heart failure, vascular changes, pulmonary edema, pleural effusion, lung opacity, cardiac changes, pericardial effusion \\
$g$6 & 32 & 4 & pleural scarring, interstitial lung disease, bronchiectasis, ild pattern \\
$g$7 & 51 & 7 & aspiration, consolidation, pneumonia, alveolar hemorrhage, atelectasis, lung opacity, pulmonary edema \\
$g$8 & 18 & 6 & hernia, vascular changes, mediastinal changes, aorta changes, goiter, bone changes \\

\bottomrule
\end{tabular}
\end{table*}
\begin{table}[htbp]
\centering
\caption{Conditional and NPMI group-size distributions across different numbers of groups ($G$).}
\label{tab:grouping_strategy}
\small
\setlength{\tabcolsep}{6pt}
\renewcommand{\arraystretch}{1.15}
\begin{tabular}{@{}cll@{}}
\toprule
$G$ & \textbf{Grouping} & \textbf{Distribution} \\
\midrule
\multirow{2}{*}{4} & Conditional & 31, 38, 56, 127 \\
 & NPMI & 16, 66, 74, 96 \\
\midrule
\multirow{2}{*}{6} & Conditional & 15, 28, 29, 31, 34, 115 \\
 & NPMI & 15, 18, 40, 45, 63, 71 \\
\midrule
\multirow{2}{*}{8} & Conditional & 11, 15, 21, 30, 30, 31, 38, 76 \\
 & NPMI & 16, 18, 23, 32, 33, 37, 42, 51 \\
\midrule
\multirow{2}{*}{10} & Conditional & 11, 14, 15, 15, 15, 18, 19, 21, 22, 102 \\
 & NPMI & 13, 16, 18, 22, 22, 24, 29, 31, 32, 45 \\
\bottomrule
\end{tabular}
\end{table}

\subsection{Grouping strategy and relation granularity}
To assess training-run variability, we evaluate NPMI, Conditional, and Random grouping over three independent training seeds. For Random grouping, labels are independently reassigned for each seed while preserving the group-size distribution of NPMI. We also conduct a controlled relation-unit ablation by comparing the default region--finding-level NPMI with a finding-level variant. The latter aggregates the regional labels into 28 finding-level variables to compute NPMI and assigns each finding's group to all corresponding regional targets, while retaining all other architectural and training settings. Table~\ref{tab:grouping_seeds} reports the mean and standard deviation across the three seeds.

\begin{table}[t]
\centering
\caption{Grouping strategy and relation-granularity results over three training seeds on the MIMIC-CXR test split. Values are mean $\pm$ standard deviation.}
\label{tab:grouping_seeds}
\small
\setlength{\tabcolsep}{4pt}
\renewcommand{\arraystretch}{1.1}
\begin{tabular}{@{}lcc@{}}
\toprule
Configuration & AUROC & AUPRC \\
\midrule
NPMI               & $0.8436 \pm 0.0023$ & $0.2031 \pm 0.0021$ \\
Conditional        & $0.8417 \pm 0.0014$ & $0.2002 \pm 0.0027$ \\
Random             & $0.8426 \pm 0.0033$ & $0.1992 \pm 0.0030$ \\
\midrule
Finding-level NPMI & $0.8404 \pm 0.0022$ & $0.2024 \pm 0.0033$ \\
\bottomrule
\end{tabular}
\end{table}

NPMI achieves the highest mean AUROC and AUPRC among the three grouping strategies. The relatively close results across grouping criteria indicate limited sensitivity to the specific group partition within the evaluated setting. Region--finding-level NPMI also yields higher AUROC than finding-level NPMI across all three seeds, with comparable AUPRC. This controlled comparison supports \modelname{}'s design choice to preserve anatomical specificity in relation construction.

\subsection{Sensitivity to \texorpdfstring{$G$}{G} and \texorpdfstring{$K$}{K}}
Table~\ref{tab:gk_sensitivity} examines the sensitivity of \modelname{} to the number of groups $G$ and prompt--expert candidates $K$.
With $K=5$ fixed, $G=8$ provides the best overall performance among the evaluated multi-group configurations.
With $G=8$ fixed, performance remains similar across nearby values of $K$.
We therefore use $G=8$ and $K=5$ as the default configuration, providing a favorable balance between predictive performance and model complexity.

\begin{table}[htbp]
\centering
\caption{Sensitivity to the numbers of co-occurrence groups and experts on the MIMIC-CXR validation split. Params. denotes the number of trainable parameters.}
\label{tab:gk_sensitivity}
\small
\setlength{\tabcolsep}{4pt}
\renewcommand{\arraystretch}{1.12}
\begin{tabular}{@{}ccc@{\hspace{10pt}}ccc@{}}
\toprule
\multicolumn{3}{c}{Varying $G$ ($K=5$)}
&
\multicolumn{3}{c}{Varying $K$ ($G=8$)}
\\
\cmidrule(lr){1-3}\cmidrule(lr){4-6}
$G$ & Params & AUROC
&
$K$ & Params & AUROC
\\
\midrule
4  & 1.73M & 0.8605 & 1 & 0.75M & 0.8588 \\
6  & 1.91M & 0.8648 & 3 & 1.42M & 0.8652 \\
\textbf{8} & \textbf{2.09M} & \textbf{0.8678}
&
\textbf{5} & \textbf{2.09M} & \textbf{0.8678} \\
10 & 2.27M & 0.8655 & 7 & 2.76M & 0.8664 \\
12 & 2.44M & 0.8620 & 9 & 3.43M & 0.8649 \\
\bottomrule
\end{tabular}
\end{table}

\section{Training and Evaluation Details}
\label{apd:second}

\subsection{Training and baseline implementation}
All classification models are trained on the same 252 region--finding label space for 20 epochs with a batch size of 512 and an input resolution of $224\times224$.

LP and FFT hyperparameters were selected independently on the validation set. For LP, the backbone is frozen and only the classifier is optimized with BCE and AdamW using a learning rate of $1\times10^{-3}$, no weight decay, and cosine decay with one warm-up epoch. For FFT, the full backbone is updated with BCE and AdamW using learning rates of $1\times10^{-5}$ for the backbone and $1\times10^{-4}$ for the classifier, weight decay of 0.05, gradient clipping at 1.0, and the same schedule.

The prompt-based baselines (VPT-Shallow, VPT-Deep, and ML-VPT) and \modelname{} use EVA-X as the frozen backbone and share the same optimization protocol: ASL ($\gamma^+=0$, $\gamma^-=2$), AdamW, a OneCycle learning-rate schedule, a learning rate of $5\times10^{-4}$, and weight decay of $1\times10^{-2}$. VPT-Shallow and VPT-Deep use 40 prompt tokens, while ML-VPT follows its original prompt architecture.

For the phrase-grounding and zero-shot CXR baselines, we use the released checkpoints and model-specific preprocessing pipelines from their official implementations~\citep{mueller2024chex,medgrounder2025,tiu2022chexzero,park2025radzero,lai2024carzero}.

\subsection{Evaluation-set exclusion and patient overlap}
\label{apd:patient_overlap}
MS-CXR and CI Gold evaluation images are excluded from the 252-label training split by image identifier. Under this protocol, 712 of 851 MS-CXR evaluation patients (83.7\%) and 335 of 500 CI Gold evaluation patients (67.0\%) have at least one other study in the training split.

We compute Jaccard similarity between positive report-derived finding sets for all distinct same-patient evaluation--training study pairs with at least one positive finding, averaging first within each patient and then across patients. The resulting similarities are 0.247 for MS-CXR and 0.145 for CI Gold.

Table~\ref{tab:patient_disjoint} reports performance for the subset of evaluation patients with no study in the 252-label training split. On these subsets, \modelname{} yields higher point estimates than ML-VPT for localization on both datasets and for CI Gold region--finding classification.

\begin{table}[htbp]
\centering
\caption{Performance on evaluation patients with no study in the 252-label training split. The task-specific subsets with valid annotations include 206 images from 136 MS-CXR patients and 98 images from 98 CI Gold patients for localization, and 149 images from 149 CI Gold patients over 58 evaluable labels for classification.}
\label{tab:patient_disjoint}
\small
\setlength{\tabcolsep}{3pt}
\renewcommand{\arraystretch}{1.08}
\begin{tabular}{@{}lllcc@{}}
\toprule
Dataset & Task & Metric & ML-VPT & CoRE \\
\midrule
MS-CXR & Localization & P.~Hit & 0.8766 & 0.8979 \\
CI Gold & Localization & P.~Hit & 0.8580 & 0.9063 \\
\midrule
\multirow{3}{*}{CI Gold} & \multirow{3}{*}{Classification}
& AUROC & 0.8594 & 0.8882 \\
& & AUPRC & 0.8247 & 0.8645 \\
& & Macro F1 & 0.4083 & 0.4506 \\
\bottomrule
\end{tabular}
\end{table}

\section{Additional Classification Analyses}
\label{apd:classification}

\subsection{Label support and rare-label analysis}
Of the 252 region--finding labels in the MIMIC-CXR test split with CI Silver annotations, one has no positive test examples, and 59 have fewer than ten, including 26 with fewer than five. Macro AUROC and AUPRC are computed only over labels with both positive and negative examples, yielding 251 labels for the point estimates. The same criterion is applied within each patient-level bootstrap resample.

To assess whether the comparison is driven by the sparsest labels, we reevaluate the same predictions on the 193 labels with at least ten positive examples (Table~\ref{tab:rare_label_subset}).

\begin{table}[htbp]
\centering
\caption{Classification performance for the 193 labels with at least ten positive examples in the MIMIC-CXR test split.}
\label{tab:rare_label_subset}
\small
\setlength{\tabcolsep}{6pt}
\renewcommand{\arraystretch}{1.1}
\begin{tabular}{@{}lcc@{}}
\toprule
Model & AUROC & AUPRC \\
\midrule
ML-VPT & 0.8238 & 0.2235 \\
CoRE   & \textbf{0.8366} & \textbf{0.2355} \\
\bottomrule
\end{tabular}
\end{table}

Using 10,000 paired patient-level bootstrap resamples, the \modelname{}--ML-VPT differences on this subset are $+0.0128$ in AUROC (95\% CI: $[0.0083, 0.0177]$) and $+0.0120$ in AUPRC (95\% CI: $[0.0020, 0.0203]$). Both intervals exclude zero, indicating that the paired differences remain positive after excluding the sparsest labels and accounting for patient-level test-sample variation. Because the full and restricted label sets differ in composition and prevalence, their absolute metrics are not directly comparable.

\subsection{Region--finding classification on CI Gold}
Because the CI Silver labels are automatically generated from radiology reports and detected anatomical regions, we complement the main evaluation with 500 CI Gold images with manually validated and corrected object--attribute annotations.

After mapping the Gold annotations to the same 252 region--finding label space, unannotated pairs are excluded rather than treated as negative, leaving 83 labels with both positive and negative examples. Table~\ref{tab:gold_classification} reports macro AUROC, macro AUPRC, and macro F1 at a fixed threshold of 0.5.

\begin{table}[htbp]
\centering
\caption{Region--finding classification on CI Gold. Results are macro-averaged over 83 labels with both positive and negative examples.}
\label{tab:gold_classification}
\small
\setlength{\tabcolsep}{5pt}
\renewcommand{\arraystretch}{1.1}
\begin{tabular}{@{}lccc@{}}
\toprule
Model & AUROC & AUPRC & F1@0.5 \\
\midrule
ML-VPT & 0.8561 & 0.8323 & 0.3507 \\
CoRE   & \textbf{0.8806} & \textbf{0.8637} & \textbf{0.3904} \\
\bottomrule
\end{tabular}
\end{table}

CoRE achieves higher values than ML-VPT for all three metrics. Because this evaluation uses a different annotation mask and an 83-label subset, its absolute metrics are not directly comparable to those reported for the MIMIC-CXR test split with CI Silver annotations in Table~\ref{tab:main_results}.

\subsection{Finding-level transfer and external label mappings}

For external finding-level evaluation, we use the findings shared between the \modelname{} label space and each target dataset. The mappings are fixed before evaluation and applied identically to all methods. For VinDr-CXR, we map `atelectasis', `consolidation', `lung opacity', `pleural effusion', `pneumothorax', and `pneumonia' to their identically named targets; `cardiac changes' to `Cardiomegaly'; `interstitial lung disease' to `ILD'; and `mass/nodule' to `Nodule/Mass'. For CheXpert, we map `atelectasis', `lung opacity', `consolidation', `pneumonia', `pneumothorax', and `pleural effusion' to their identically named targets; `cardiac changes' to `Cardiomegaly'; and `pulmonary edema' to `Edema'. For \modelname{}, each finding-level score is obtained by taking the maximum logit across its corresponding regional outputs.

As an additional supervised reference, we fully fine-tune EVA-X directly on the same 28 findings using the same MIMIC-CXR training split. Table~\ref{tab:fft28_external} compares this model with \modelname{} on the two external datasets.

\begin{table}[htbp]
\centering
\caption{External finding-level performance of EVA-X FFT-28 and CoRE.}
\label{tab:fft28_external}
\small
\setlength{\tabcolsep}{4pt}
\renewcommand{\arraystretch}{1.1}
\begin{tabular}{@{}llcc@{}}
\toprule
Dataset & Model & AUROC & AUPRC \\
\midrule
\multirow{2}{*}{VinDr-CXR}
& EVA-X FFT-28 & 0.9346 & 0.4863 \\
& CoRE          & 0.9313 & 0.4978 \\
\midrule
\multirow{2}{*}{CheXpert}
& EVA-X FFT-28 & 0.9111 & 0.6833 \\
& CoRE          & 0.9125 & 0.6653 \\
\bottomrule
\end{tabular}
\end{table}

EVA-X FFT-28 achieves higher VinDr-CXR AUROC and CheXpert AUPRC, whereas \modelname{} achieves higher VinDr-CXR AUPRC and CheXpert AUROC. Neither supervision strategy is consistently superior across both datasets and metrics. These results indicate that \modelname{}'s fine-grained predictions over 252 region--finding labels can also be aggregated for conventional finding-level evaluation without target-domain adaptation.

\section{Spatial Attribution and Localization Metrics}\label{apd:third}

For all classification methods, target-specific patch attribution is computed using the same protocol.
We use the patch representations at the output of the first layer-normalization module in the final ViT block.

For region--finding target $i$ and patch $m$, we compute a gradient--activation attribution score at the selected patch representation~\citep{ancona2018towards}:
\begin{equation}
S_{i,m}(x)
=
\left[
\mathbf{t}_m(x)^{\top}
\frac{\partial \ell_i(x)}
{\partial \mathbf{t}_m(x)}
\right]_{+},
\end{equation}
where $\ell_i(x)$ is the target logit, $\mathbf{t}_m(x)$ is the representation of patch $m$ at the selected layer, and $[\cdot]_{+}$ retains positive attribution.

For each positive region--finding instance, let $B_i$ denote the union of ground-truth bounding boxes associated with the corresponding anatomical region.
P.~Hit follows the pointing-game criterion~\citep{zhang2016topdown}.
We identify the patch with the highest attribution,
\begin{equation}
m_i^{\star}
=
\arg\max_m S_{i,m}(x),
\end{equation}
and count a hit when the center of that patch falls within $B_i$ in the original image coordinates:
\begin{equation}
\mathrm{Hit}_i
=
\mathbb{1}
\left[
c(P_{m_i^{\star}})\in B_i
\right],
\end{equation}
where $P_m$ denotes the image-space extent of patch $m$ and $c(P_m)$ its center.
P.~Hit is obtained by averaging $\mathrm{Hit}_i$ over the evaluated instances.

Under hierarchy propagation, attribution for a parent target is computed from the output that attains the propagated maximum, whether the parent or a descendant, and its peak is evaluated against the parent ground-truth box. We do not maximize P.~Hit separately across hierarchical outputs.

For finding-level localization, let $\mathcal{I}(f)$ denote the set of region--finding labels associated with finding $f$.
We select the region--finding output with the maximum logit,
\begin{equation}
i_f^{\star}
=
\arg\max_{i\in\mathcal{I}(f)} \ell_i(x),
\end{equation}
and compute patch attribution with respect to $\ell_{i_f^{\star}}(x)$.
A hit is counted when the center of the highest-attribution patch falls within the union of ground-truth boxes annotated for finding $f$.
Following the standard EVA-X preprocessing, images for the classification models are resized and center-cropped to $224\times224$. To maintain a common evaluation set across methods, two apical annotations whose boxes became empty after this transformation are excluded, leaving 1,160 of 1,162 MS-CXR phrase-level annotations. Although this affected only two annotations, center cropping may reduce coverage of anatomy near image boundaries.


\section{Qualitative Error Case Analysis}\label{fourth}

\begin{figure}[htbp]
    \centering
    \includegraphics[width=\columnwidth]{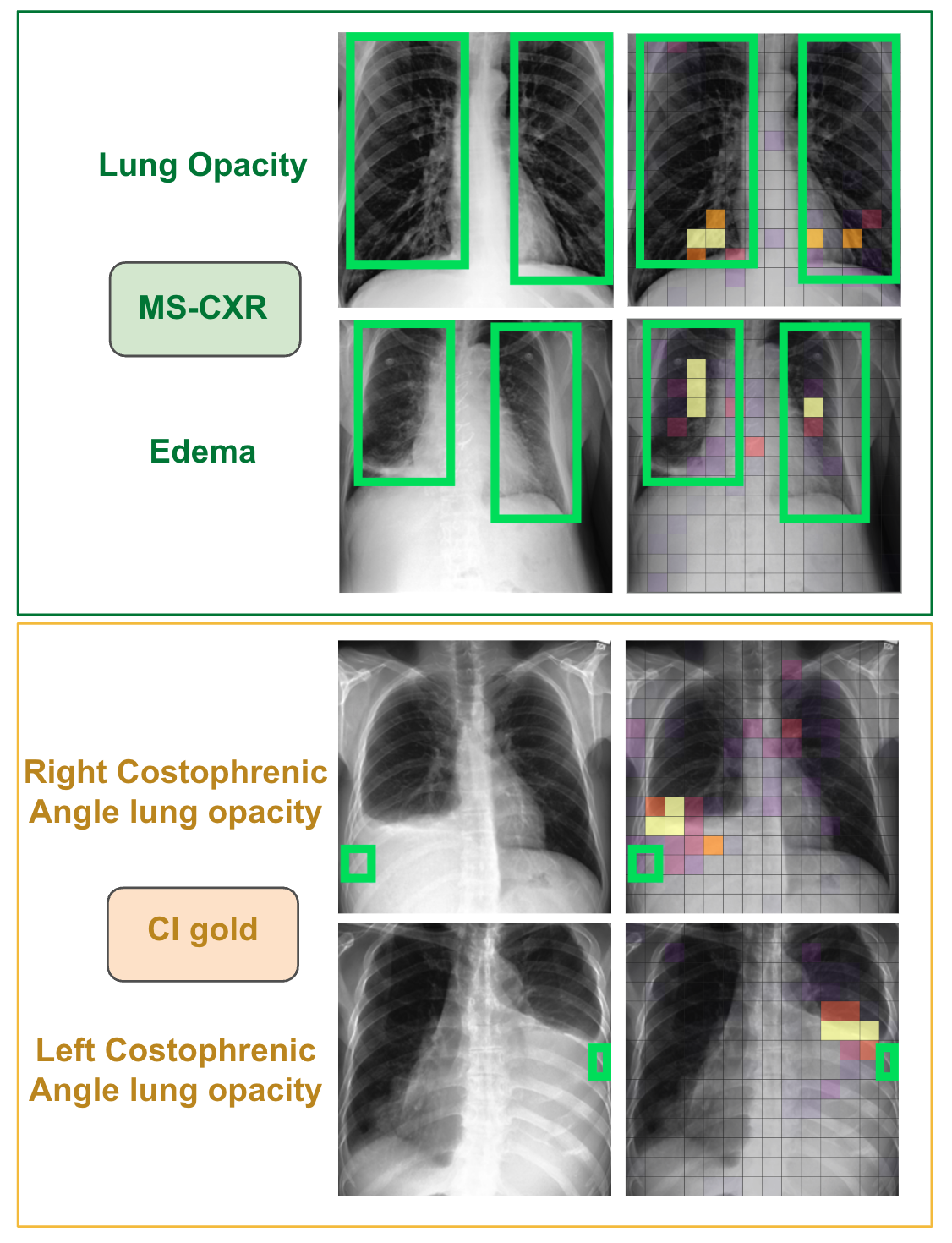}
    \caption{\footnotesize
    \textbf{Representative localization error cases of \modelname{}.}
    Each row shows the ground-truth annotation and corresponding patch-saliency map.
    The first two rows show spatially extended lung opacity and edema from MS-CXR, and the last two show right and left costophrenic angle--lung opacity from the CI gold standard.
    }
    \label{fig:error}
\end{figure}

Figure~\ref{fig:error} examines representative error cases observed in the spatial alignment evaluation.

For spatially extended abnormalities, saliency from the corresponding region--finding outputs can remain within anatomically relevant areas while covering only part of the annotated finding extent, as observed for lung opacity and edema.
This reflects a limitation of the current formulation: \modelname{} is optimized for region--finding classification rather than dense lesion delineation and therefore does not explicitly enforce coverage of the full abnormality.

Other apparent errors arise from the strict correspondence between the predicted region--finding target and its annotated box.
In the right and left costophrenic-angle examples, the highest-saliency response falls outside the target box and is therefore counted as an error by P.~Hit, yet the activation remains concentrated on an adjacent area exhibiting the same abnormal finding.
These examples suggest that some P.~Hit errors may reflect spatial discrepancies with the annotated target rather than entirely irrelevant model evidence.

\end{document}